\documentclass[english,aps, preprint]{revtex4}

\usepackage{graphicx}

\makeatletter

\usepackage{babel}
\makeatother
\begin{document}

\preprint{Preprint}

\title{Measuring the Temperature of a Mesoscopic Electron System by means of Single Electron Statistics}

\author{Enrico Prati}

\affiliation{Laboratorio Nazionale Materiali e Dispositivi per la Microelettronica,
Consiglio Nazionale delle Ricerche - Istituto Nazionale per la Fisica della Materia, Via Olivetti 2, I-20041
Agrate Brianza, Italy}

\email{enrico.prati@cnr.it}

\author{Matteo Belli}

\affiliation{Laboratorio Nazionale Materiali e Dispositivi per la Microelettronica,
Consiglio Nazionale delle Ricerche - Istituto Nazionale per la Fisica della Materia, Via Olivetti 2, I-20041
Agrate Brianza, Italy}

\author{Marco Fanciulli}

\affiliation{Laboratorio Nazionale Materiali e Dispositivi per la Microelettronica,
Consiglio Nazionale delle Ricerche - Istituto Nazionale per la Fisica della Materia, Via Olivetti 2, I-20041
Agrate Brianza, Italy}

\affiliation{Dipartimento di Scienza dei Materiali, Universita degli Studi Milano-Bicocca, 20125 Milano, Italy}

\author{Giorgio Ferrari}

\affiliation{Dipartimento di Elettronica e Informazione, Politecnico di Milano,
P.za Leonardo da Vinci 32, I-20133 Milano, Italy}

\begin{abstract}
We measure the temperature of a mesoscopic system consisting of an ultra-dilute two dimensional electron gas at the $Si/SiO_2$ interface in a metal-oxide-semiconductor field effect transistor (MOSFET) by means of the capture and emission of an electron in a point defect close to the interface. Contrarily to previous reports, we show that the capture and emission by point defects in Si n-MOSFETs can be temperature dependent down to 800 mK. As the finite quantum grand canonical ensemble applies, the time domain charge fluctuation in the defect is used to define the temperature of the few electron gas in the channel.
\end{abstract}


\maketitle

We create a mesoscopic system made of a quantum gas of few electrons in thermal and particle exchange with a point defect capable of containing one or two electrons, for which a generalized temperature is determined by virtue of the finite quantum grand partition ensemble.


Nanoelectronics \cite{Rogge06,Rogge08,Prati08i} and the study of mesoscopic systems in solid state devices at low temperature \cite{Giazotto06} consider ensembles of small numbers of interacting particles. A single donor atom in a semiconductor \cite{Rogge06,Prati08i}, defects \cite{Prati08}, and quantum dots \cite{Sanquer00} may contain one or two localized electrons. These electrons are in thermal contact with a thermal bath consisting of a two or three dimensional electron system confined in the leads or in a nanometric channel. There, the total number of electrons may be of the order of tens up to several hundreds. Temperature is defined  by the inverse of the partial derivative of the energy with respect to the entropy, at a fixed volume \cite{Landau}. Unlike in macroscopic system, for which the number of electrons in the reservoir is infinite and conventional statistical physics holds, the temperature of a mesoscopic system consisting of few electrons confined in a nanostructure is defined by virtue of a time domain extension of the grand canonical ensemble in the limit of small finite $N$ \cite{Prati09ii}. 
A system constituted by a refrigerated ultra-dilute two dimensional electron gas at the Si/SiO$_2$ interface in a Metal-Oxide-Semiconductor Field-Effect-Transistor (MOSFET) and a zero dimensional defect located at the interface realizes an experimental condition well described by the finite quantum grand canonical ensemble.\cite{Prati09ii} The point defect can exchange one electron with the two dimensional system by a thermally assisted tunneling.

Capture and emission of a single electron by defects are monitored by the current fluctuactions through the channel of the MOSFET. Such effect goes under the name of Random Telegraph Signal (RTS) \cite{Ralls86,Palma97,Prati06,Prati08}, because of the peculiar bi-stable current values flowing randomly as a function of time. 
In conventional theory of RTS, the same temperature is attributed to the lattice and to the electrons. \cite{Palma97} The temperature has been previously extracted from the dependence of $\tau_c /\tau_e$ from a static magnetic field in the case of an infinite thermal bath of electrons in the channel.\cite{Prati06} 


We report on the experimental determination of the generalized temperature of a mesoscopic ensemble of few electrons, for which the generalized temperature is defined by means of the finite quantum grand canonical ensemble.\cite{Prati09ii}
First, we demonstrate that the detailed balance principle holds below 10 K by measuring the temperature dependence of the occupancy probability of defect at the $Si/SiO_2$ interface, contrarily to previous reported experiments \cite{Scofield00,Buehler04}. Next, we experimentally determine the generalized temperature of the system constituted by the electrons confined in a defect and the ultra-dilute 2-dimensional electron gas (2DEG).


The experiment was carried in a n-channel MOS device having
a length of 180 nm and a width of 280 nm, gate oxide thickness of
3.5 nm and a threshold voltage at 4.2 K of $V_T$= 490 mV.\cite{Cappelletti04}
The drain current $I_{DS}$ was measured by a transimpedance amplifier with bandwidth from dc to about 10 kHz. The electronics is powered by independent batteries to avoid power-line
pick ups and interferences, while the sample was controlled by a 12 bits data acquisition board. The drain and gate voltages can be set with a resolution better than 50 $\mu V$.
This setup allowed us to characterize traps with a characteristic time down to 100 $\mu$s. 
The sample was immersed in liquid $^{3}He$. We measured the sample at a temperature below 10 K down to the base temperature of 290 mK. 
In order to deal with a small number of identical particles, an electron system confined at the interface between Si and the SiO$_2$ has been obtained by cooling a MOSFET polarized close to the threshold voltage $V_T$ down to the nominal temperature of $T=290$ mK. The number of electrons in the 2DEG has been roughly estimated by the charge stored in the MOS capacitor:  $Q= \epsilon_{ox} WL/t_{ox} (V_{G}-V_T)$, where $W$,$L$ are the width and length of the transistor and $t_{ox}$ the oxide thickness. By working few mV above the threshold voltage, a number of electrons in the order $N\approx16$ has been obtained.
In confirmation of the low number of electrons, the charge transport at gate voltages close to the threshold $V_{T}$ exhibits Coulomb blockade characteristics (Figure 1) typical of sequential tunneling through a few electron quantum dot starting from $N_{QD}=0$.\cite{Beenaker91} The current flows through an electrostatic island placed between the source and the drain.  
We observe that transport is affected by RTS at small bias and close to the threshold. We model the system constituted by the quantum dot formed in the channel and the interface defect as depicted in Figure 2. The defect exchanges one electron with the channel partially formed below the gate at one side of the electron island. 
In contrast with previously reported results \cite{Scofield00}, the random telegraph signal in the n-MOSFET was found to be temperature dependent far below 10 K. In Figure 3 the exponential trend of the ratio of the measured emission time $\tau_e$ and capture time $\tau_c$ is reported between 0.6 K and 1.75 K. The ratio $\tau_e / \tau_c$ depends exponentially on the inverse of the nominal temperature $T$. We conclude that the principle of detailed balance holds in the temperature range considered. The heating of the electron gas was minimized by applying a small bias in the proximity of the threshold voltage. The current is of the order of few nA. This aspect represents the major difference between our experiment and those reported in the Ref. \cite{Scofield00} where the current was of the order of $\mu A$.
We measured the effective generalized temperature of the electron gas by studying the B field dependence of the ratio between the occupancies in the time domain.
The trap is a paramagnetic center with one unpaired electron, capable of capturing and emitting a second electron (1$\rightarrow$2). Such conclusion is provided by the measurement of the characteristic times $\tau_{c}$ (capture) and $\tau_{e}$ (emission) as a function of the
gate voltage $V_{G}$ and of the magnetic field up to 3 T at the nominal temperature of 310 mK.\cite{Prati06}
The ratio between the probabilities of realization of one and two particles occupancy of a mesoscopic system constituted by a cell in thermal and particle exchange with a few electron gas of $N$ identical particles at the generalized temperature $\Theta$ is described by the Equation 
\begin{equation}
\frac{p(1)}{p(2)}= 2 e^{\beta \left( E_T+\Delta E_L+\Delta U- \mu (N) \left( 1+ \frac{3}{2N} \right) \right)}  
\end{equation}
where $\beta=(k_B \Theta)^{-1}$, $\mu$ is the chemical potential, $\Delta U$ is the charging energy when two electrons occupy the point defect, $\Delta E_L$ is the energy gain of the lattice when the second electron is captured, and $N$ is the number of electrons of the dilute gas.\cite{Prati09ii} The expression Eq. 1 follows from the experimental details of the mechanisms involved in low temperature RTS.
At low temperature the 2DES is weakly coupled with the crystal \cite{Kivinen03}. Phonons are involved in the emission and capture of one electron from the defect and the relaxation of the crystal implies a change of energy of $S \hbar \omega$ where $S$ is the Huang-Rhys factor and $\omega$ is the average phonon frequency in the configuration coordinate picture \cite{Goguenheim90}. We identify the energy gain of the lattice when an electron is captured with the Huang-Rhys factor $\Delta E_L = S \hbar \omega$. Similarly to the charging energy $\Delta U$ of the electrostatic potential of the defect, $\Delta E_L$ can be accounted for a negative variation of the electron system Gibbs free energy. 
Under the ergodic assumption, the ratio between the average occupation times equals the ratio between the occupation probabilities, so the experimentally accessible characteristic times provide information on the statistics via: $p(1)/p(2)=\tau_c / \tau_e$.
The ratio $\tau_c/\tau_e$ experimentally observed as a function of the magnetic field $B$ is reported in Figure 4. The deviation from a pure exponential trend with respect to the $B$ field reveals that both the singlet and the triplet states are accessible at sufficiently high magnetic field.\cite{Prati06} The contribution of the energy shift from the single electron occupancy to the two electrons occupancy of the point defect $\Delta_{12}=\Delta E_L+\Delta U$ is temperature independent. Therefore the magnetic field dependence of $\tau_c/\tau_e$ is similar to that presented in Ref. \cite{Prati06}:
\begin{equation}
\frac{\tau_c}{\tau_e}=\frac{\gamma e^{\beta \left( E_T+\Delta_{12}- \mu (N) \left( 1+ \frac{3}{2N} \right) \right) }\left(\cosh\beta\left\langle E_{Z}\right\rangle +\cosh\beta\frac{\delta}{2}\right)}{1+e^{-\beta \Delta_{ST} }\left(1+2 \cosh\beta E_{Z}^{T}\right)}
\end{equation}
where $\gamma$ is the degeneracy factor, $\left\langle E_{Z}\right\rangle =\frac{E_{Z}^{T}+E_{Z}^{C}}{2}$ is the average value of the two Zeeman splittings of the channel and of the trap respectively, $\delta=\left|E_{Z}^{T}-E_{Z}^{C}\right|$ and $\Delta_{ST}$ is the energy gap between the singlet and the triplet energy levels in the doubly occupied defect.
The generalized temperature for such system extracted from the fitting is $T_{fit}=0.80(1)$ K. $\Delta_{ST}$=150(5) $\mu$eV is due to the different charging energy caused by the different orbital symmetry of the electron wavefunction for the singlet and the triplet states. The difference between the generalized electron temperature and the lattice temperature which is in thermal equilibrium with the $^{3}He$ at 310 mK is consistent with other reported values in similar systems.
To conclude, we experimentally determined the generalized temperature of a system of few electrons at cryogenic temperature. We have shown that below 10 K the capture and the emission of a point defect may be thermally activated and we discussed the role of phonons. We realized such mesoscopic system in a n-MOSFET, where a point defect exchanges one electron with the 2DES at the Si/SiO$_2$ interface. The generalized temperature is extracted from the probability of occupation of a mesoscopic system constituted by a point defect capable of containing either 1 or 2 electrons in thermal contact with a dilute electron system, in the limit of finite $N$, with the help of a static magnetic field. This method provides an estimate of the temperature of a system consisting of a localized electron in thermal exchange with a small number of electrons confined in two dimensions.

\section{Figures}

\begin{figure}[hbt]
\begin{center}\includegraphics[scale=0.7, angle=0]{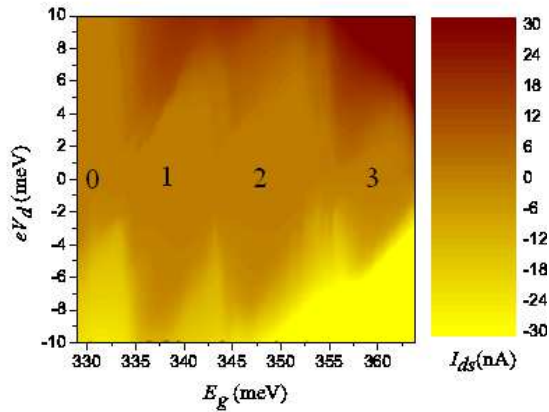}
\caption{The current stability diagram at 300 mK. The energy $E_G$ consists of the gate voltage $V_G$ times the coupling constant $\alpha=1.428$. The n-MOSFET manifests a quantum dot behaviour, which reveals the formation of an electron island in the Silicon channel starting from the $N_{QD}=0$ state.}
\label{fig:Figure1}
\end{center}
\end{figure}

\begin{figure}[hbt]
\begin{center}\includegraphics[scale=0.5, angle=0]{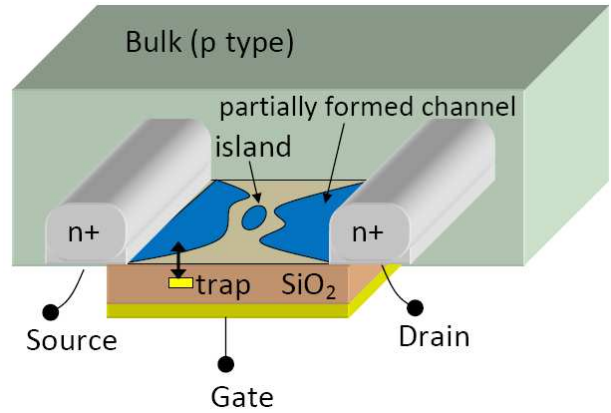}
\caption{Model of a n-MOSFET at cryogenic temperature with a paramegnatic trap close to the $Si/SiO_2$ interface:  the current flows by sequential tunelling through the island formed in the Silicon channel. The current fluctuates between two levels according to the charge occupation of the interface defect which captures and emits an electron at one of the two separated two-dimensional systems between the central island and the corresponding contact.
}
\label{fig:Figure2}
\end{center}
\end{figure}

\begin{figure}[hbt]
\begin{center}\includegraphics[scale=0.8, angle=0]{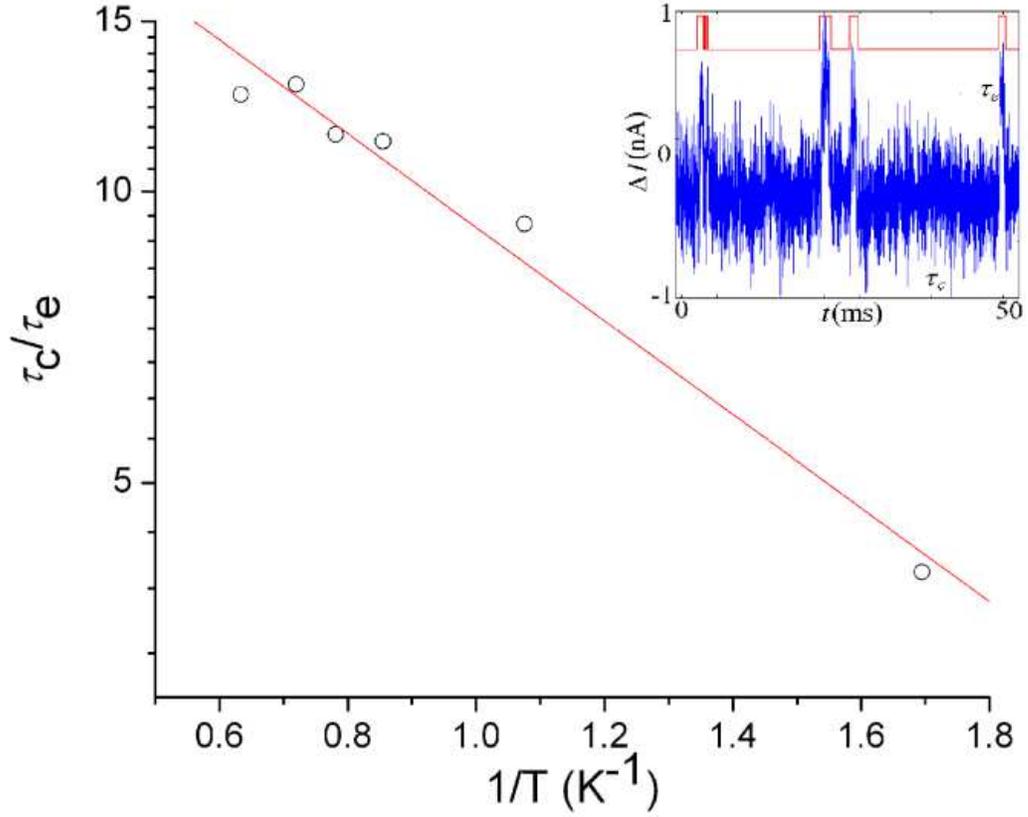}
\caption{The relative occupation $\tau_e / \tau_c$ versus the inverse of the nominal temperature $1/T$. We demonstrate that in our sample the detailed balance holds down to the nominal value of 600 mK. The sample was operated at $V_G$=495.4 mV and $V_d$=-9 mV and it carried a current of about 15 nA. Inlet: the current switching as function of time at the lowest temperature. The current shift $\Delta I$ is 0.9 nA.
}
\label{fig:Figure3}
\end{center}
\end{figure}

\begin{figure}[hbt]
\begin{center}\includegraphics[scale=0.8, angle=0]{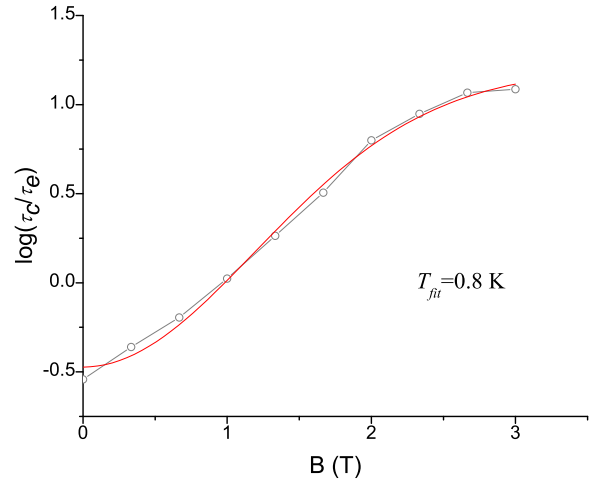}
\caption{Magnetic field dependence of $\tau_c/\tau_e$. The deviation from a pure exponential trend reveals the formation of a triplet state at sufficiently high field. The generalized temperature extracted from the experimental data is $T_{fit}=0.80(1)$ K. }
\label{fig:Figure4}
\end{center}
\end{figure}

\end{document}